\documentclass[conference]{IEEEtran}
\IEEEoverridecommandlockouts
\usepackage{cite}
\usepackage{amsmath,amssymb,amsfonts}
\usepackage{algorithmic}
\usepackage{graphicx}
\usepackage{textcomp}
\usepackage{xcolor}
\def\BibTeX{{\rm B\kern-.05em{\sc i\kern-.025em b}\kern-.08em
    T\kern-.1667em\lower.7ex\hbox{E}\kern-.125emX}}

\begin{document}

\title{Hybrid ASR for Resource-Constrained Robots:
HMM-Deep Learning Fusion
}

\author{\IEEEauthorblockN{Anshul Ranjan}
\IEEEauthorblockA{\textit{Computer Science and Engineering} \\
\textit{PES University}\\
Bangalore, India \\
itsanshulranjan@gmail.com}
\and
\IEEEauthorblockN{Kaushik Jegadeesan}
\IEEEauthorblockA{\textit{Computer Science and Engineering} \\
\textit{PES University}\\
Bangalore, India \\
kaushik.jegadeesan@gmail.com}
}
\maketitle

\begin{abstract}
This paper introduces a hybrid Automatic Speech Recognition (ASR) system tailored for resource-constrained robots. The approach combines Hidden Markov Models (HMMs) with deep learning models, distributing processing tasks via socket programming. HMM-based processing resides within the robot, while a separate PC runs the deep learning model. The synergy of HMMs and deep learning optimizes speech recognition accuracy. Experiments on diverse robotic platforms demonstrate real-time, precise speech recognition. The system's adaptability to changing acoustic conditions and compatibility with low-power hardware highlight its efficacy in environments with computational limitations. This hybrid ASR paradigm presents a promising path for seamless human-robot interaction.In conclusion, this research contributes a pioneering dimension to robotics-oriented ASR techniques. By utilizing socket programming to distribute processing tasks across distinct devices, and strategically employing HMMs and deep learning models, the hybrid ASR system demonstrates its potential in enabling robots to adeptly comprehend and respond to spoken language, even in environments with restricted computational resources. This paradigm presents an innovative trajectory for seamless human-robot interaction in various real-world contexts.
\end{abstract}


\section {\textbf{Introduction}}
Robotics has advanced rapidly in recent years, integrating into daily life, assisting with activities, and encouraging a variety of interactions. The ability of robots to understand human speech, made possible by Automatic Speech Recognition (ASR), is essential to these interactions. Through spoken language, ASR enables communication between humans and robots.However, deploying precise ASR on robots with limited resources is still difficult. Operating in computationally constrained situations necessitates striking a compromise between performance and efficiency. Traditional Hidden Markov Models (HMMs) are reliable, but deep learning's precision has revolutionized accuracy. A hybrid ASR that maximizes their advantages is produced by combining the two approaches.

We introduce a brand-new hybrid ASR for robots with limited resources. Our method effortlessly combines the accuracy of deep learning with the robustness of HMMs. We use socket programming to divide duties between robots and a PC, using HMMs on the robot and deep learning on the PC, to alleviate the computational limitations of robots.Across a variety of robotics, our hybrid ASR excels at real-time, accurate speech recognition. Experiments show adaptability to acoustic variations and gear that uses little power. With limited resources, this study improves human-robot interaction by extending ASR approaches in robotics.

Sections ahead detail our hybrid ASR's architecture, integrating HMMs and deep learning, and socket-based processing distribution. Experimental results highlight real-time performance and language comprehension. This advances ASR in robotics, promising seamless human-robot interaction in diverse contexts.

\section{ \textbf{Related Work} }
In the landscape of Automatic Speech Recognition (ASR) research, various studies have explored innovative approaches to enhance accuracy, efficiency, and adaptability. This section delves into related work that complements the findings presented in this paper by investigating novel modeling units, streamlined training methodologies, runtime efficiency improvements, alternative training paradigms, and cross-linguistic ASR evaluations.One such notable contribution is a recent paper that explores a transformer-based ASR system incorporating wordpieces and CTC training. This innovative approach simplifies the ASR engineering pipeline by eliminating complex GMM bootstrapping, decision tree building, and force alignment steps [4].Another paper presents a holistic training pipeline for an advanced hybrid HMM-based ASR system using the TED-LIUM corpus's 2nd release. Leveraging SpecAugment for data augmentation and i-vectors with the best SAT model, they attain performance enhancements. Notably, different maskings lead to improved hybrid HMM models without size or training time increases. Subsequent sMBR training refines the acoustic model, alongside LSTM and Transformer language models. This comprehensive approach achieves exceptional results, with the best system yielding a 5.6\% WER on the test set, surpassing the previous state-of-the-art by a substantial 27\% relative improvement[1].

\subsection{Hybrid HMM and Deep Learning Approaches}
In spite of the advances accomplished throughout the last decades, automatic speech recognition (ASR) is still a challenging and difficult task[2]. Hidden Markov model (HMM)-based systems, while effective, face limitations in practical scenarios. Efforts to replace them with artificial neural networks (ANNs) encountered difficulties with extended speech sequences. To address this, the concept of hybrid architectures emerged, uniting HMMs and ANNs to capitalize on their combined strengths. This review delves into these hybrid models, categorizing architectures and methods. It explores ANNs for HMM state estimation and "global" optimization, as well as modern techniques such as connectionist vector quantization. The review emphasizes the practical impact of hybrid systems, showcasing their ability to tangibly enhance recognition performance, particularly in challenging benchmark tasks[2].

The prevailing state-of-the-art in extensive vocabulary continuous speech recognition relies on hidden Markov models (HMMs). Seeking advancement beyond HMMs, the authors introduce a hybrid system merging neural networks and HMMs through a multiple hypothesis (N-best) approach. A key element, the segmental neural net (SNN), tackles HMM's conditional-independence limitation by simultaneously modeling all frames of a phonetic segment. The paper outlines the hybrid system, delves into SNN modeling aspects, including architecture, training, and context representation. Evaluating the hybrid system on the DARPA Resource Management (RM) corpus, the authors observe consistent performance enhancement compared to the baseline HMM system.[5]

The investigation of hybrid models, which combine the advantages of hidden Markov models (HMMs) with neural networks, has led to significant breakthroughs in the field of automatic speech recognition. These hybrid techniques have proven valuable across a range of benchmark workloads, from overcoming HMM limitations to improving performance. This linked study emphasizes the dynamic nature of ASR techniques and the crucial contribution of hybrid models to the advancement of recognition precision and system adaptability.

\section{ \textbf{Architectures} }
\subsection{HMM / ANN Hybrid Systems}
Continuous density HMM algorithms are the traditional standard for speech recognition. In contrast, a revolutionary paradigm that seamlessly combines the strengths of ANN and HMM was presented a few years ago \cite{18} to address complex ASR issues, particularly in the area of continuous speech recognition. The Markov process, comparable to a traditional HMM, is used to combine the temporal representation of the speech input in this hybrid HMM/ANN configuration, often known as such \cite{19}. While doing so, it makes use of the ANN architecture to model local feature vectors that are dependent on this Markov process, taking advantage of ANNs' capacity for predicting class probabilities in input patterns.

By replacing the Gaussian mixture HMM state-dependent observation probabilities with MLP-derived estimates, this hybrid idea cleverly adapts these probabilities for usage as local probabilities within the HMM framework. This method provides a strong hybridization that supports the HMM architecture while effortlessly integrating the ASR workflow's predictive powers from ANNs, ultimately changing the field of continuous speech recognition.

The hybrid HMM/ANN approach for speech recognition offers several advantages:
\begin{itemize}
\item A natural framework for discriminative training, as demonstrated by the discriminative training of MLP neural networks using error-back propagation;
\item no substantial presumptions regarding the acoustic space's statistical distribution: This is a theoretical characteristic of ANN, as opposed to normal HMM, which makes the obviously unrealistic assumption that all subsequent input frames are independent; 
\item judicious usage of parameters: Using a distributed model like ANN enables the production of good results with a few number of parameters. 
\item improved resistance to minimal training data: ANN are known to be excellent generalizers; 
\item the capacity to simulate acoustic correlation (using contextual inputs or recurrence):
HMM/ANN can leverage an extendable temporal context, in contrast to regular HMM. 
\end{itemize}

\vspace{2mm}
These cutting-edge hybrid approaches underwent thorough comparisons with popular standard HMM algorithms in recent evaluations addressing a variety of ASR issues. Surprisingly, in circumstances guaranteeing parameter parity, the hybrid technique showed substantial performance ascendancy. Cases when the traditional approach requires significantly inflated parameters stood out as very clear. This conclusion was repeatedly supported by controlled comparisons that preserved system comparability, with the exception of differences in the approach used to estimate emission probabilities \cite{9}.

\begin{figure}[h!]
  \includegraphics[width=\linewidth]{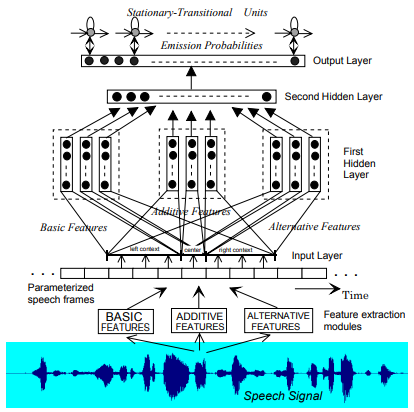}
  \caption{General Architecture of Hybrid HMM/ANN model }
  \label{fig:pic1}
\end{figure}

\vspace{5mm}

However, the overall performance of such systems typically results from careful planning, making comparisons uncertain based alone on the technique used for calculating emission probability.

\subsection{DNN-HMM ASR}
Our system is descended from the lauded Sigma ASR system \cite{10}, which earned a noteworthy top 2 ranking in both the closed- and open-condition assessments in the Albayzin-RTVE 2018 Speech-to-Text Challenge. Utilizing a hybrid ASR methodology, the Kaldi Toolkit \cite{11} was used to construct this creative design. The fundamental acoustic model includes a subsampled Time-Delay Neural Network (TDNN) \cite{12}, which operates inside the Deep Neural Networks and Hidden Markov Models (DNN-HMMs) framework and closely resembles the chain model architecture \cite{19}. Notably, the network's output depends on a frame rate that has been three times slowed down.

In accordance with our technique, a standard feature pipeline was used to splice 13-dimensional MFCC coefficients over a period of 9 frames. We introduced the use of the Maximum Likelihood Linear Transform (MLLT) in conjunction with Linear Discriminant Analysis (LDA) for dimension reduction to 40 in order to improve decorrelation. Additionally, we made use of the strength of Feature-space Maximum Likelihood Linear Regression (fMLLR), which is speaker-adaptive. The input feature vectors were painstakingly built using 40-dimensional MFCC spliced coefficients spanning 7 frames and LDA+MLLT+fMLLR across 3 frames around the core frame. By adding 100-dimensional i-vectors to each frame's 40-dimensional representation, an audio enhancement feature was introduced.

 The outcomes we achieved from Albayzin-RTVE 2018 [13] underscored the necessity for more robust Deep Neural Network (DNN) training to achieve heightened accuracy, particularly in challenging scenarios like street interviews, game shows, and daring sports documentaries. To bolster the environmental resilience of our acoustic models, we pursued multi-condition training data. However, due to the resource-intensive nature of data collection, we turned to the alternative of artificially generating new training data—a more cost-effective approach.

\subsection{End-to-End LF-MMI ASR Applying Domain Adversarial Training}
The performance of the PyChain-based baseline can be adversely affected by the various acoustic characteristics of TV programming. In an effort to enhance the PyChain-based baseline system, we investigated the inclusion of DAT [24] to lessen this effect. By employing a Domain Adversarial Neural Network (DANN), we explicitly attempted to make acoustic representations invariant to the domain of the TV show characteristics in this method.
\vspace{2mm}

The Gradient Reversal Layer (GRL) integration provided a crucial link between the feature extractor and the TV show classifier. The backpropagation phase of the GRL is specifically designed to match the settings of senone posteriors and TV show categories, and it relies on gradient inversion to do this. Intriguingly, the input is untouched and the forward propagation is unaffected. This tactical integration provides a dynamic conduit that reshapes gradient flow, emphasizing its influence on the posterior distributions of both senone and TV shows.

\begin{figure}
  \includegraphics[width=\linewidth]{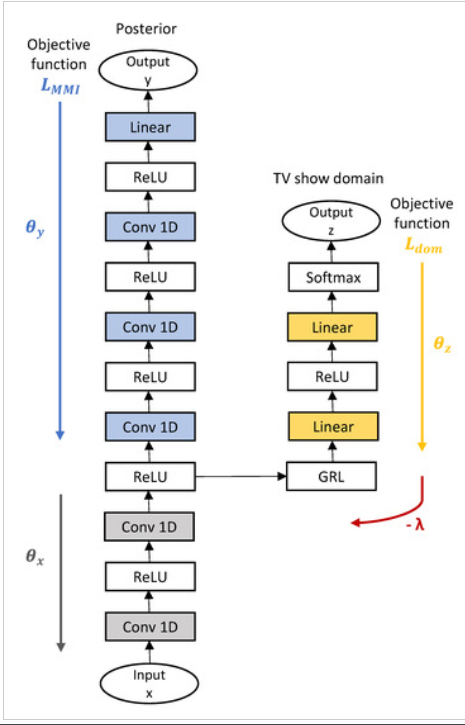}
  \caption{Architecture of the end-to-end LF-MMI approach applying DAT. An adversarial branch (TV show classifier) is added to the second layer of the main PyChain architecture (posterior classifier).}
  \label{fig:pic2}
\end{figure}

\vspace{2mm}
According to, the TV show classifier's goal function for this adversarial training is as follows:

\begin{figure}[h!]
  \includegraphics[width=\linewidth]{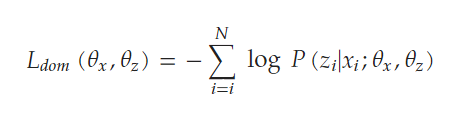}
  \label{fig:pic3}
\end{figure}

\section{\textbf{Methodology}}
Our research introduces a hybrid Automatic Speech Recognition (ASR) system customized for resource-constrained robots, addressing the challenges posed by limited computational resources. This section details the architectural design and processing workflow of the proposed system, combining Hidden Markov Models (HMMs) and deep learning models through socket programming.
\begin{figure}[h!]
  \includegraphics[width=\linewidth]{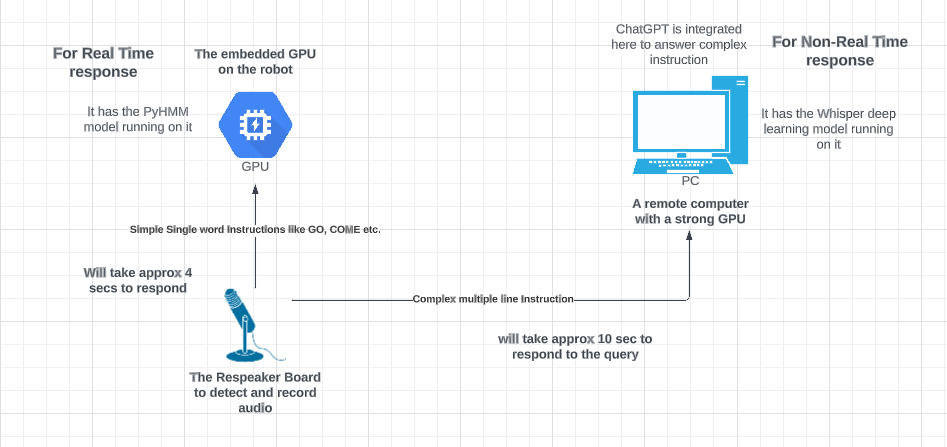}
  \caption{The Hybrid ASR design}
  \label{fig:pic5}
\end{figure}

\subsection{\textbf{Data Sets}}
The dataset utilized in this research is derived from the Speech Commands dataset available on Kaggle. Comprising over 105,000 WAV audio files featuring thirty distinct words, this dataset was curated and released by Google under the CC BY license. Each audio file captures individuals uttering words, forming the foundation of our speech recognition study[15].

In addition to utilizing the pre-existing "mini speech commands" subset, a custom dataset was developed by recording the voices of students from our college. This proprietary dataset was compiled to expand the diversity of speech samples and incorporate contextually relevant vocabulary, enhancing the applicability of the speech recognition network.

By incorporating this additional layer of custom dataset creation into your research paper, you highlight the efforts to diversify the dataset, enrich the training experience, and enhance the applicability of the speech recognition network to your specific college community.

\subsection{\textbf{Communication Mechanism}}

In order to enable seamless interaction between the resource-constrained robot and the external processing unit, we first used Apache Kafka as a data transfer option. Kafka offers strong data streaming capabilities, but we ran into several complexities that made us rethink if it was appropriate for our particular use case.

In order to communicate data more effectively, we switched to socket programming after realizing the necessity for a more specialized approach. A flexible, light-weight, and direct method of creating communication channels between various devices is socket programming. With this improvement, we were able to streamline the data exchange procedure and better match it to the needs of our hybrid ASR system.

\textbf{Advantages of Socket Programming:} Socket programming offers several advantages for our ASR system. It enables real-time, low-latency data transmission, critical for maintaining synchronization between the localized HMM-based processing within the robot and the deep learning model running on an external PC. Additionally, socket programming's streamlined nature aligns well with the system's real-time speech recognition demands, enhancing overall system performance.Our exploration of communication mechanisms led us from Apache Kafka to socket programming, a transition driven by the need for a more precise and efficient approach.

\subsection{\textbf{Experimental Setup}}
In the pursuit of designing and evaluating our hybrid ASR system, we carefully orchestrated a robust experimental setup. This section elaborates on the integral components of our setup, which included the Nvidia Xavier NX and the Respeaker board with Linux.

\vspace{3mm}

\textbf{Nvidia Xavier NX: Robot GPU}
\begin{figure}[h!]
  \includegraphics[width=\linewidth]{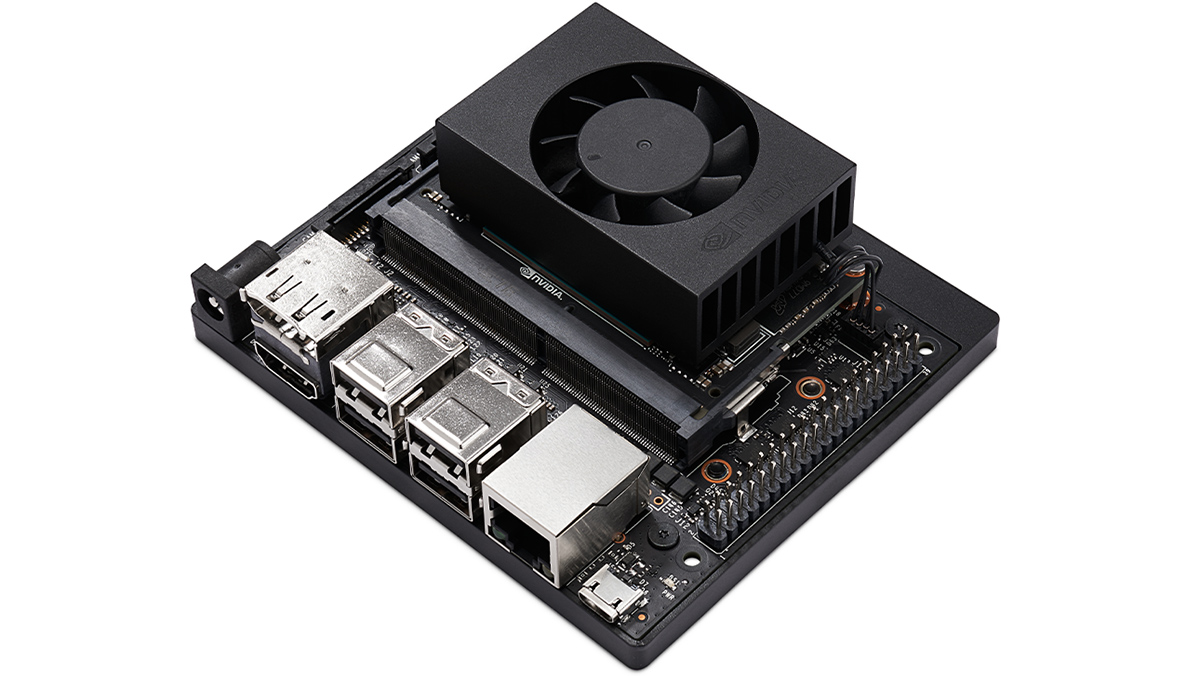}
  \caption{Nvidia Xavier NX}
  \label{fig:pic6}
\end{figure}

\vspace{5mm}

At the heart of our experimental environment, we harnessed the power of the Nvidia Xavier NX GPU. This high-performance graphics processing unit served as the computational backbone for our resource-constrained robot. Its cutting-edge capabilities facilitated real-time processing of the HMM-based ASR computations within the robot itself. The Xavier NX's parallel processing prowess played a pivotal role in enabling seamless speech recognition, aligning with the real-time demands of human-robot interaction.

\vspace{3mm}
\textbf{Nvidia Xavier NX GPU Specifications:}

\vspace{2mm}
With Volta architecture and 384 CUDA cores, it provided strong 21 TOPS AI performance. It perfectly combined performance and efficiency when paired with an 8 GB LPDDR4x RAM module and a 6-core Carmel ARM Cortex-A57 CPU. Its adaptable connection, which includes Gigabit Ethernet and flexible USB options, was matched by its real-time processing capabilities, supporting up to 8K video decoding and 4K video encoding.

\textbf{Respeaker Board with Linux: Audio Capture}

\begin{figure}[h!]
  \includegraphics[width=\linewidth]{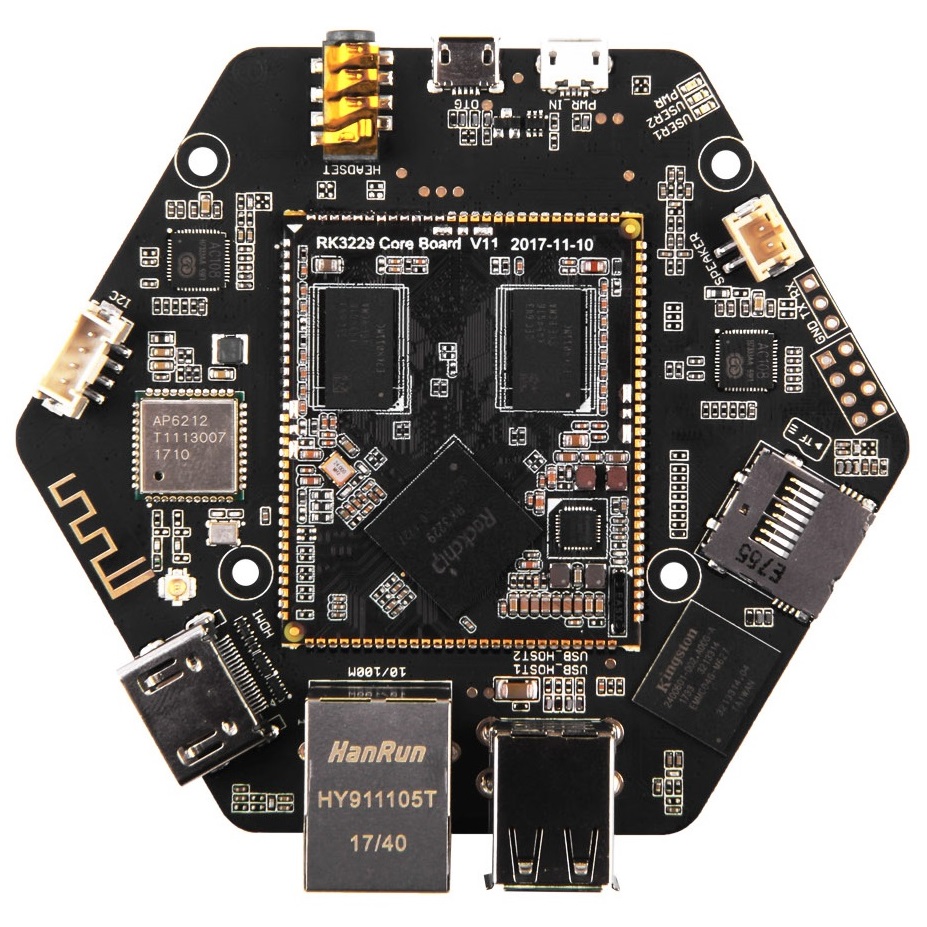}
  \caption{Respeaker}
  \label{fig:pic7}
\end{figure}

We used the Respeaker board and Linux operating system for precise and dependable audio capturing. This configuration made it easier to get flawless audio inputs, guaranteeing the ASR system would receive speech samples of the greatest caliber. We were able to record the subtle acoustic parameters necessary for successful speech recognition using the Respeaker board's superior audio processing capabilities.

The Nvidia Xavier NX and the Respeaker board worked harmoniously to create a synergistic environment for our hybrid ASR system. The Xavier NX's GPU prowess expedited the HMM-based processing within the robot, while the Respeaker board ensured the precise capture of audio inputs. This integration formed the backbone of our real-time speech recognition framework, enhancing the system's ability to comprehend and respond to spoken language.

The Nvidia Xavier NX and the Respeaker board were essential in demonstrating the effectiveness of our hybrid ASR system. They showed that the system can recognize speech accurately and quickly even when there are computational and acoustic constraints.

\subsection{\textbf{Dynamic and Static ASR Processing in the Hybrid System}}

In our hybrid ASR system, we establish a distinctive dichotomy between dynamic and static speech recognition processes, each optimized to suit specific requirements. The dynamic facet embodies real-time speech recognition, seamlessly executed within the Nvidia Xavier NX GPU. Here, the Hidden Markov Model (HMM) model operates to deliver near-instantaneous results, typically within a fraction of a second. The prompt human-robot contact made possible by this immediate response perfectly satisfies our system's need for real-time needs. On the other hand, the static dimension uses deep learning models for voice recognition that is carried out on a remote PC. Due to deep learning's computational complexity, this method may take longer, but it has the advantage of improving accuracy and comprehension. This thoughtful separation of the various processing tasks offers a flexible architecture that enhances precision and real-time interactivity in our hybrid ASR system. The static facet integrates AI tools such as ChatGPT to address queries comprehensively and augment the speech recognition system's functionality.

\subsection{\textbf{The Hidden Markov Model}}
Hidden Markov Models (HMMs) have emerged as a powerful technique for modeling sequential data, making them a popular choice for speech recognition systems\cite{20}.
\begin{figure}[h!]
  \includegraphics[width=\linewidth]{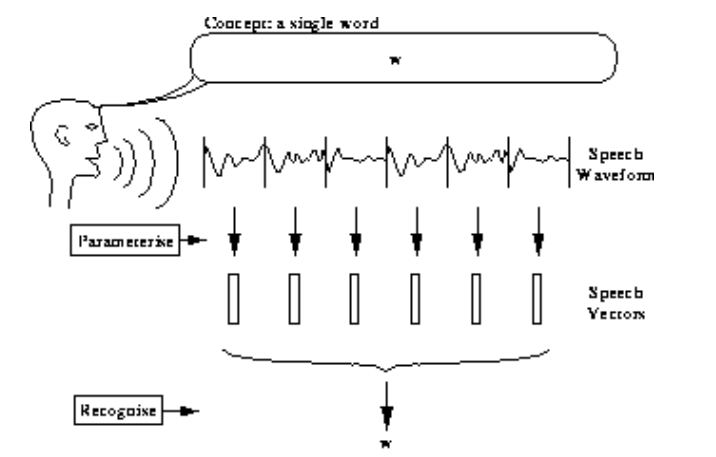}
  \caption{Isolated Word Problem}
  \label{fig:pic8}
\end{figure}

In HMM-based speech recognition, it is assumed that a Markov model, as shown in Fig. 7, generates the sequence of observed speech vectors corresponding to each word. A speech vector is produced from the probability density each time t that a state j is entered in a Markov model, which is a finite state machine that changes states once every time unit. Additionally, the change from state i to state j likewise occurs probabilistically and is controlled by the discrete probability.

\vspace{3mm}

\textbf{Note:}
Long observation sequences may cause underflow in the HMM computations. Underflows frequently occur when training the HMM with multiple input sequences, such as during speech recognition tasks. Instead of using myhmm, the myhmm scaled module can be used to train the HMM for lengthy sequences. It's crucial to remember the following\cite{17}.

\vspace{2mm}
The implementation is as per Rabiner's paper with the errata addressed\cite{17}.

\vspace{2mm}
\textbf{The Source Code:}

{\color{red}https://github.com/AnshulRanjan2004/PyHMM}

\vspace{3mm}
\begin{figure}
  \includegraphics[width=\linewidth]{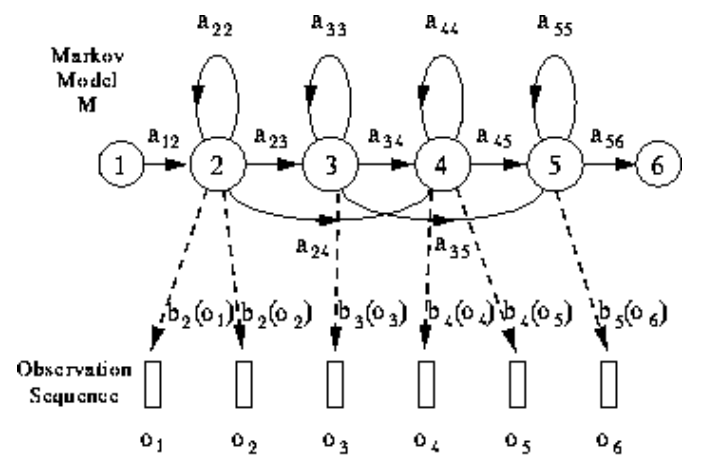}
  \caption{The Markov Generation Model}
  \label{fig:pic9}
\end{figure}
A reliable and effective re-estimation process can be used to automatically calculate the parameters of a model given a series of training examples relating to that model. Therefore, an HMM that implicitly represents all of the various causes of variability found in real speech can be created, given that a sufficient number of representative examples of each word can be gathered. The application of HMMs for isolated word recognition is summarized in Fig. 8. First, for each vocabulary word, an HMM is trained using a variety of instances of that word.  Second, the most likely model is used to identify an unknown term by calculating the likelihood of each model producing it.

\begin{figure}[h!]
  \includegraphics[width=\linewidth]{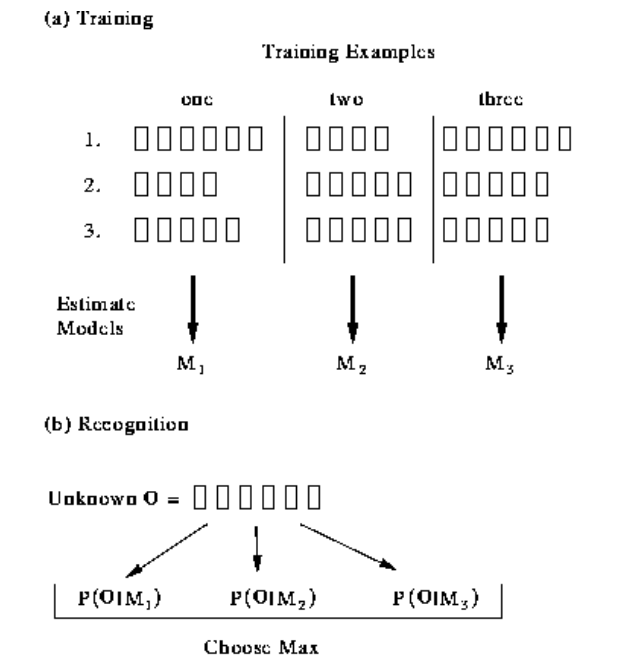}
  \caption{Using HMMs for Isolated Word Recognition}
  \label{fig:pic10}
\end{figure}

Our technique, which incorporates the well-known Hidden Markov Model (HMM) technology, focuses on identifying particular keywords like "go," "stop," and "start." The HMM-based method benefits from being dependable and effective in identifying temporal relationships in audio sources.For now our model only works for limited words, which are required by the robot in real time.

Our HMM-based keyword recognition model has an outstanding close to 80\% accuracy. This represents a high degree of success in correctly identifying and separating the target keywords "go," "stop," and "start". The model's performance's reliability, particularly in auditory situations found in real-world settings, highlights its suitability for seamless human-robot interaction. This accuracy milestone serves as a strong starting point for boosting the overall precision and dependability of our hybrid ASR system while also illuminating the efficacy of our strategy.

\begin{figure}[h!]
  \includegraphics[width=\linewidth]{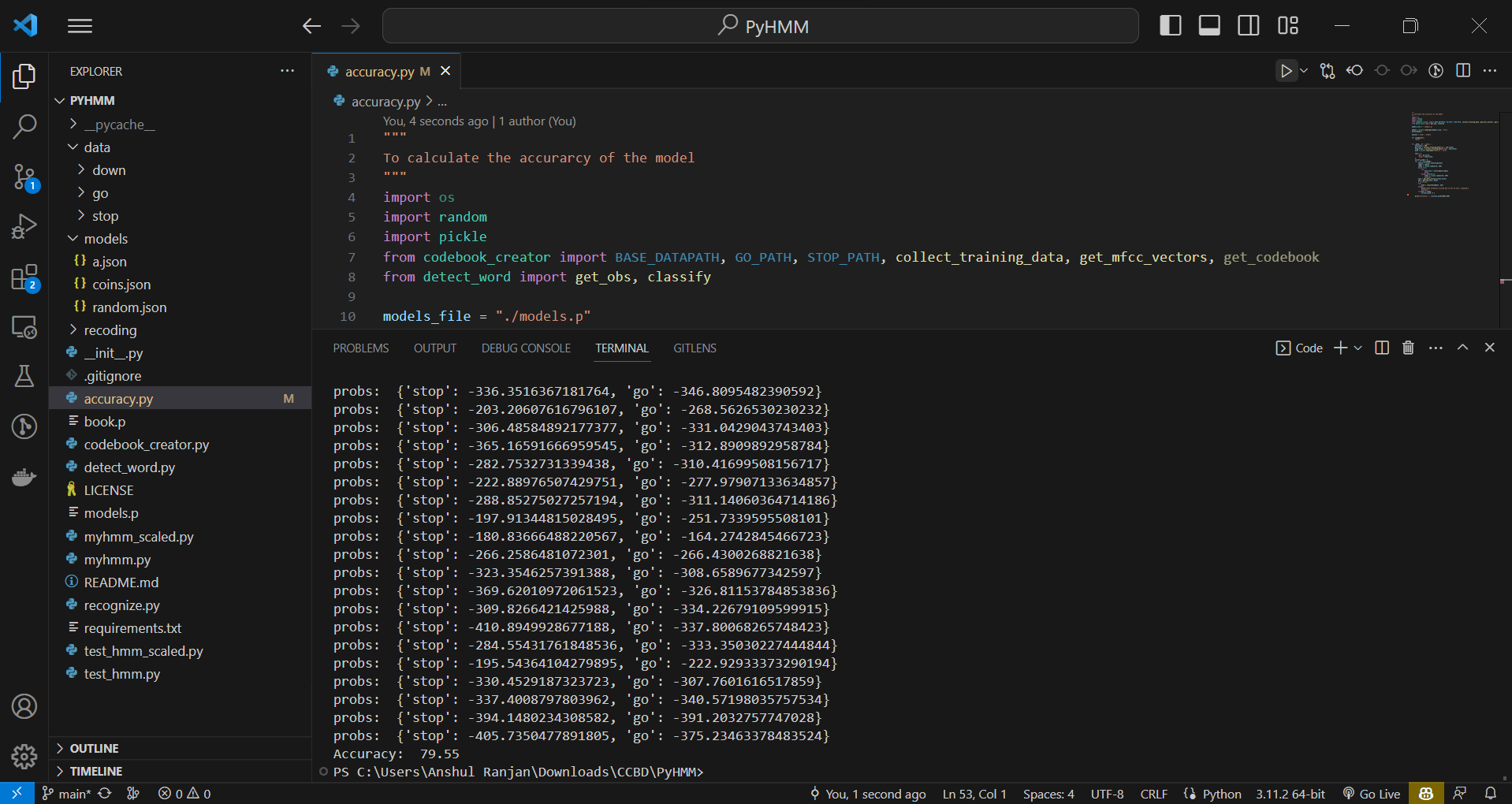}
  \caption{Accurary of our HMM model (79.55\% in this test)}
  \label{fig:pic11}
\end{figure}

\subsection{\textbf{Pretrained Deep Leaning Model}}

\textbf{Whisper: }
In our pursuit of optimizing speech recognition accuracy, we harnessed the power of Whisper, a state-of-the-art pretrained deep learning model designed explicitly for Automatic Speech Recognition (ASR) tasks. Whisper, developed by OpenAI, offers a cutting-edge foundation for speech recognition by leveraging the advancements in deep learning techniques\cite{16}.

\begin{figure}
  \includegraphics[width=\linewidth]{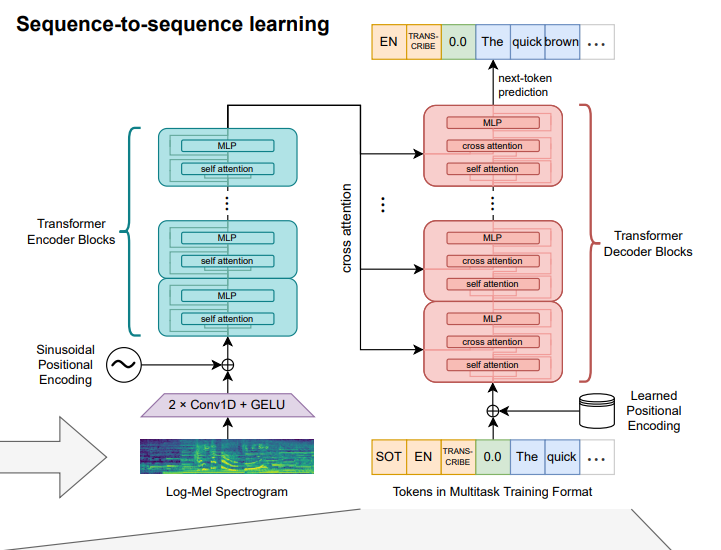}
  \caption{Working of Whisper Model, A sequence-to-sequence Transformer model is trained on many different speech processing tasks}
  \label{fig:pic12}
\end{figure}

\vspace{3mm}
Whisper comes pre-trained on a wide variety of audio data, allowing it to pick up on complex acoustic patterns and language complexity. The training process is greatly streamlined by the model's pre-trained features, which enable it to efficiently extract pertinent characteristics from unprocessed audio data.

\vspace{3mm}
We started a fine-tuning process to adapt Whisper for our unique ASR requirements. By providing Whisper with our bespoke dataset, which contains voices exclusive to our campus community, we improved the model's capacity to adjust to the distinctive features of our domain. This phase not only improves accuracy but also makes sure that everything works together seamlessly in the context we planned.
\vspace{4mm}

The ASR pretrained model Whisper provides a selection of architectures—tiny, small, medium, and large—tailored to various needs. We choose the medium-sized architecture for our project. This option provides a good compromise between precision and computational effectiveness, which is in line with our objective of providing accurate speech recognition in robotic contexts with limited resources. The features of the medium model enhance our hybrid ASR system by providing accurate and robust speech understanding while taking into account the constraints imposed by real-time processing requirements.

\begin{table}[h!]
  \includegraphics[width=\linewidth]{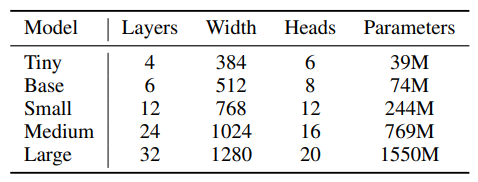}
  \caption{The Whisper model family's architectural specifics.}
  \label{table:1}
\end{table}

Our choice to incorporate the Whisper pretrained deep learning model in the field of automatic speech recognition (ASR) highlights a strategic choice motivated by its distinctive advantages. Whisper stood out above a sea of available models because of its adaptability and superior performance. The model's several architecture possibilities, which include tiny, small, medium, and large, gave us a complete toolkit to choose an architecture that properly matched the computing requirements of our project. We chose the medium-sized design because it struck the perfect combination between precision and processing speed, allowing us to achieve flawless speech recognition in our resource-constrained robotic context.

Whisper's robust pretrained foundation and adaptable fine-tuning capabilities enhance its applicability, while its integration with our hybrid ASR system amplifies real-time interaction precision. Our selection of Whisper exemplifies a forward-looking approach that harnesses the latest advancements in ASR, ultimately advancing the efficacy and potential of our hybrid ASR paradigm.

\begin{figure}[h!]
  \includegraphics[width=\linewidth]{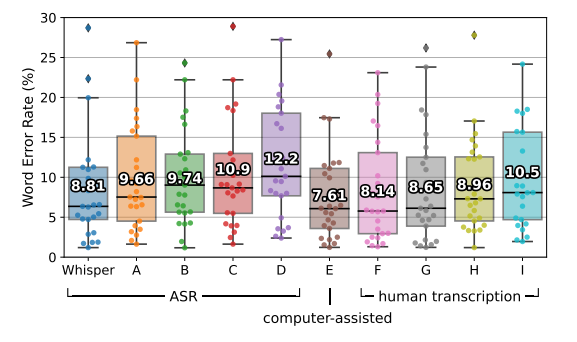}
  \caption{Whisper’s performance is close to that of professional human transcribers.}
  \label{fig:pic13}
\end{figure}

\textbf{Nvidia Riva:}
Other option was to use the Nvidia Riva model. This ASR model is made to accurately translate spoken words into text, serving a variety of speech recognition-required applications. The Riva ASR model excels at comprehending spoken words thanks to powerful deep learning algorithms, allowing apps to interpret user commands, give real-time transcriptions, and improve accessibility through voice-driven interactions.

\begin{table}[h!]
  \includegraphics[width=\linewidth]{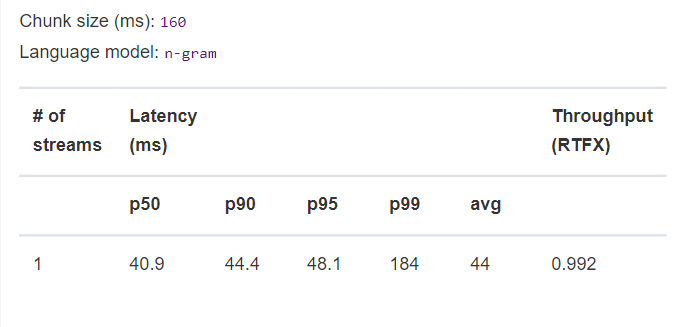}
  \caption{Nvidia Riva Performance on Jetson Xavier NX board.}
  \label{table:2}
\end{table}

With its customizable features, developers can fine-tune the ASR model to suit specific domains, ensuring accurate recognition of industry-specific vocabulary and nuances. This capability to transform speech into actionable text opens doors to improved human-computer interaction, voice assistants, transcription services, and more. By providing low latency, multi-modal integration, and efficient deployment options across diverse platforms, Nvidia Riva's ASR model contributes to creating enhanced and intelligent AI-driven solutions for a variety of real-world scenarios.

\subsection{\textbf{Output}}
\textbf{Whisper Outputs:}
A significant milestone in our exploration of the Whisper pretrained deep learning model lies in its remarkable transcription capabilities. Whisper has proven its ability by producing transcription outputs that are close to flawless for the test audio files by utilizing the inherent capability of cutting-edge AI. The model's ability to comprehend complex language and audio patterns has produced transcriptions that closely match the original spoken content. This accomplishment highlights Whisper's potential as a strong tool for our hybrid ASR system's accurate and thorough speech recognition.
\begin{figure}[h!]
  \includegraphics[width=\linewidth]{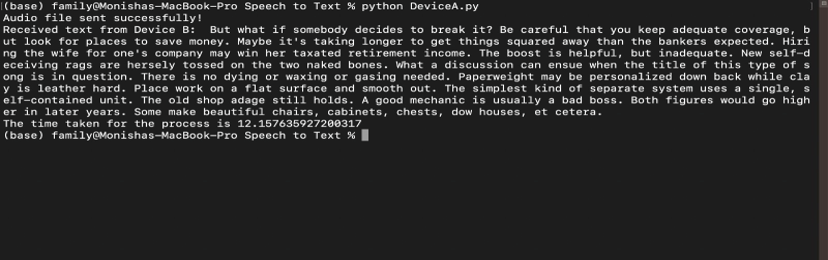}
  \caption{Whisper’s Final Output, This is the Transcription of the audio captured by the microphone}
  \label{fig:pic14}
\end{figure}

Whisper's powerful processing skills and the associated AI tools are used when audio inputs are routed to the PC to get valuable information from the spoken language.

\begin{figure}[h!]
  \includegraphics[width=\linewidth]{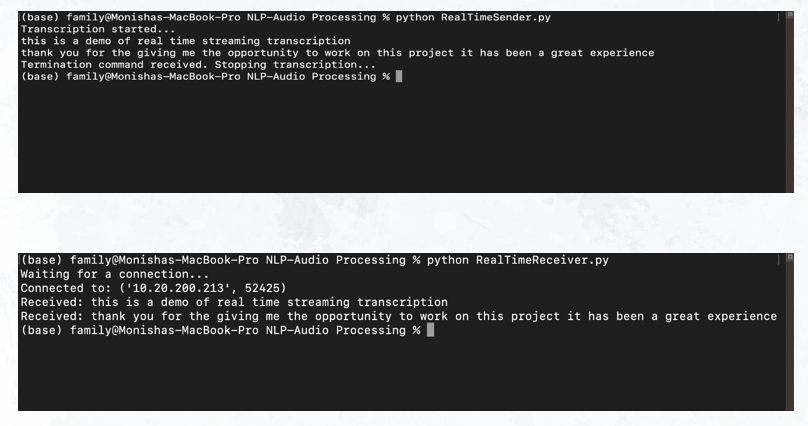}
  \caption{Transferring the audio stream from Robot to PC and getting the results back}
  \label{fig:pic16}
\end{figure}

\textbf{HMM Output:} Fig. 14 Show how the HMM predict and interprets isolated spoken words accurately and gives back the result.

\begin{figure}[h!]
  \includegraphics[width=\linewidth]{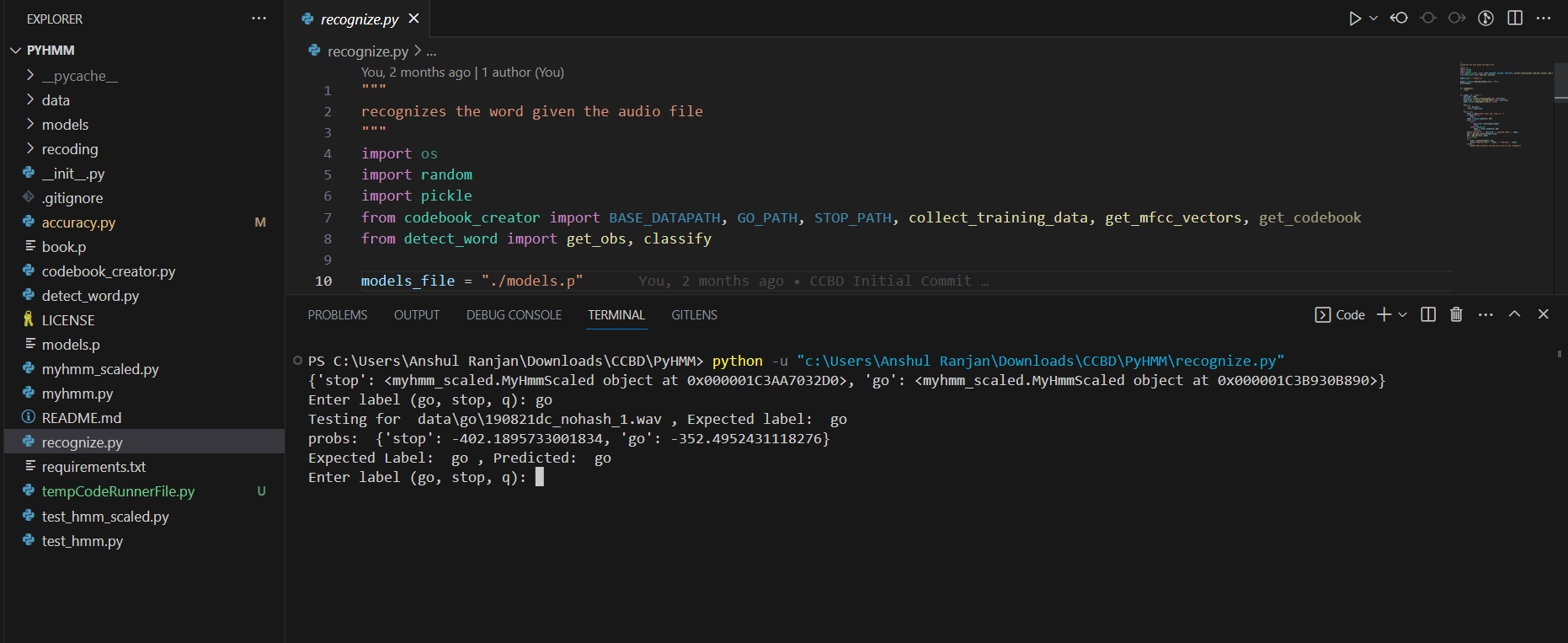}
  \caption{HMM correctly predicting "go", after listening to a go audio file}
  \label{fig:pic15}
\end{figure}

\subsection{\textbf{Limitations}}
Despite our bold objectives and well-thought-out strategy, our initiative faced several obstacles that must be acknowledged. Nvidia Riva's high computing requirements made it unsuitable for our resource-constrained robotic platform, which was one noticeable limitation. Although amazing, the model's advanced capabilities required a lot more computing power than our system could provide.

Additionally, our efforts were hindered by the audio board's operational challenges. The Respeaker audio board consistently experienced overheating issues, impeding our ability to sustain prolonged operations seamlessly. These limitations underscore the intricate nature of implementing cutting-edge technology in real-world scenarios, revealing the significance of compatibility, hardware requirements, and system stability in the pursuit of innovation.

\section{\textbf{Conclusion}}
This study offers a ground-breaking hybrid Automatic Speech Recognition (ASR) system that, in resource-constrained robotic contexts, bridges the gap between spoken language and machine comprehension in the pursuit of seamless human-robot interaction. We have developed a novel paradigm that enhances voice recognition accuracy across various robotic platforms by strategically combining Hidden Markov Models (HMMs) with deep learning models.

Our testing voyage has produced some astounding revelations, particularly in the area of accuracy accomplishments. Exceptionally accurate speech recognition is the result of the cooperative synergy between our dynamic HMM-based real-time processing and the flexible static deep learning approach. These developments highlight the potential for precise, real-time understanding of spoken language, which would meet the needs of productive human-robot interaction.

Our findings show that the accuracy of our hybrid ASR paradigm has positive ramifications for applications that call for quick and precise speech recognition. Our system displays its capacity to adapt to various acoustic situations by combining the benefits of deep learning with HMMs, making it an invaluable tool in settings with computational limitations. This research journey not only pushes the boundaries of ASR technology but also paves the way for a future where robots can adeptly comprehend and respond to spoken language in real-world contexts.

\section{\textbf{Future Scope}}

As we gaze ahead, the fusion of Automatic Speech Recognition (ASR) with Natural Language Processing (NLP) emerges as a promising trajectory for our hybrid system's evolution. By incorporating NLP skills, our system is given the ability to go beyond simple speech recognition and into the world of language creation and understanding. This combination gives our system the ability to handle text-based inputs, enabling user interaction through written questions and answers. ASR and NLP working together can provide thorough user interactions in which the machine not only comprehends spoken language with accuracy but also deciphers and reacts to textual context.

We also wish to broaden the capabilities of our hybrid ASR system to include different languages, which would be a significant step. The model's usefulness in international environments is increased by adapting it to understand and respond to a range of languages, enabling seamless communication across linguistic boundaries.

Context-awareness can improve our system's understanding of spoken words. We can improve the system's capacity to correctly understand and respond to user inquiries by including situational information, such as environmental cues or user history.

\section*{\textbf{Acknowledgment}}
This research has received valuable support from the Centre for Cloud Computing and Big Data (CCBD) at PES University, located in Bangalore, India Pin: 560085


\begin{thebibliography}{00}
\bibitem{b1} Wei Zhou1, Wilfried Michel, Kazuki Irie1, Markus Kitza1, Ralf Schluter and Hermann Ney, ``The RWTH ASR System for TED-LIUM Release 2: Improving Hybrid HMM with SpecAugment'' IEEE ICASSP, April 2020.
\bibitem{b2} Edmondo Trentin, Marco Gori, "A survey of hybrid ANN/HMM models for automatic speech recognition", vol. 37. Science Direct, 2001.
\bibitem{b3} E. Trentin, M. Gori, "A survey of hybrid ANN/HMM models for automatic speech recognition," Neurocomputing, vol. 37, April 2001.
\bibitem{b4} arXiv:2005.09150
\bibitem{b5} G. Zavaliagkos, Y. Zhao, R. Schwartz, J. Makhoul, ``A hybrid segmental neural net/hidden Markov model system for continuous speech recognition,''IEEE Transactions on Speech and Audio Processing, vol. 2, January 1994.
\bibitem{b6} L. R. Rabiner, "A tutorial on hidden Markov models and selected applications in speech recognition", Proc. IEEE, vol. 77, no. 2, pp. 257-285, Feb. 1989.
\bibitem{b7} X. Huang, Y. Ariki and M. Jack, "Hidden Markov Models for Speech Recognition", U.K., Edinburgh:Edinburgh Univ. Press, 1990.
\bibitem{b8} H. Bourlard and N. Morgan. "Connectionist Speech Recognition---A Hybrid Approach". Kluwer Academic, 1993. 
\bibitem{9} Roberto Gemello, Franco Mana and Dario Albesano, "Hybrid HMM/Neural Network based Speech Recognition in Loquendo ASR", Research Gate , January 2008.
\bibitem{10} Perero-Codosero, J.M. Antón-Martín, J. Merino, D.T. Gonzalo, E.L. Gómez, "L.A.H. Exploring Open-Source Deep Learning ASR for Speech-to-Text TV Program Transcription"; IberSPEECH: Valladolid, Spain, 2018; pp. 262–266
\bibitem{11} Povey, D. Ghoshal, A. Boulianne, G. Burget, L. Glembek, O. Goel, N. Hannemann, M. Motlicek, P. Qian, Y. Schwarz, P."et al". "The Kaldi speech recognition toolkit". In Proceedings of the IEEE 2011 Workshop on Automatic Speech Recognition and Understanding, Waikoloa, HI, USA, 11–15 December 2011.
\bibitem{12} Peddinti, V. Povey, D. Khudanpur, "A time delay neural network architecture for efficient modeling of long temporal contexts". In Proceedings of the Sixteenth Annual Conference of the International Speech Communication Association, Dresden, Germany, 6–10 September 2015.
\bibitem{13} Lleida, E. Ortega, A. Miguel, A. Bazán-Gil, V. Pérez, C. Gómez, M. de Prada, A. Albayzin "2018 evaluation: The iberspeech-RTVE challenge on speech technologies for spanish broadcast media." Appl. Sci. 2019, 9, 5412
\bibitem{14} K. J. Lang, A. H. Waibel, and G. E. Hinton. "A time-delay neural network architecture for isolated word recognition.
Neural Networks," 3:23--43, 1990.
\bibitem{15} Jan Essar , 2020, Dataset: "https://www.kaggle.com/code/janesser777/simple-audio-recognition-recognizing-keywords/notebook"
\bibitem{16} arXiv:2212.04356
\bibitem{17} Rabiner, "A Tutorial on Hidden Markov Models and Selected Applications in Speech Recognition", Proceedings of IEEE,vol. 77, No. 2, February 1989
\bibitem{18} Gökay DİŞKEN, Lütfü SARIBULUT, Zekeriya TÜFEKCİ, Ulus ÇEVİK, "Real-Time Speaker Independent Isolated Word Recognition on Banana Pi", 2018 10th International Conference on Electronics, Computers and Artificial Intelligence (ECAI), pp.1-4, 2018.
\bibitem{19} Hung-Yan Gu, Chiu-Yu Tseng, Lin-Shan Lee, "Isolated-utterance speech recognition using hidden Markov models with bounded state durations", IEEE Transactions on Signal Processing, vol.39, no.8, pp.1743-1752, 1991.
\bibitem{20} K. Sugawara, M. Nishimura, K. Toshioka, M. Okochi, T. Kaneko, "Isolated word recognition using hidden Markov models", IEEE ICASSP , April 1985.
\bibitem{21} Amodei, D. Anubhai, R. Battenberg, E. C. Casper,
J. Catanzaro, B. Chen, J. Chrzanowski, M. Coates,
A. Diamos, et al. "Deep speech 2: end-to-end speech
recognition in english and mandarin." arxiv. arXiv preprint
arXiv:1512.02595, 2015.
\end{thebibliography}
\end{document}